\documentclass[showpacs,preprintnumbers,amsmath,amssymb]{revtex4}
\usepackage{graphicx}
\usepackage{dcolumn}
\usepackage{bm}

\begin{document}
\newcommand{\ri}{{\rm i}}
\newcommand{\re}{{\rm e}}
\newcommand{\bx}{{\bf x}}
\newcommand{\br}{{\bf r}}
\newcommand{\bk}{{\bf k}}
\newcommand{\bE}{{\bf E}}
\newcommand{\bR}{{\bf R}}
\newcommand{\bn}{{\bf n}}
\newcommand{\rSi}{{\rm Si}}
\newcommand{\beps}{\mbox{\boldmath{$\epsilon$}}}
\newcommand{\rg}{{\rm g}}
\newcommand{\tr}{{\rm tr}}
\newcommand{\xmax}{x_{\rm max}}
\newcommand{\ra}{{\rm a}}
\newcommand{\rx}{{\rm x}}
\newcommand{\rs}{{\rm s}}
\newcommand{\rP}{{\rm P}}
\newcommand{\up}{\uparrow}
\newcommand{\down}{\downarrow}
\newcommand{\hc}{H_{\rm cond}}
\newcommand{\kb}{k_{\rm B}}
\newcommand{\cI}{{\cal I}}
\newcommand{\cE}{{\cal E}}
\newcommand{\cC}{{\cal C}}
\newcommand{\Ubs}{U_{\rm BS}}
\sloppy

\title{A Quantitative Measure of Interference} 
\author{Daniel Braun and Bertrand Georgeot}
\affiliation{Laboratoire de Physique Th\'eorique, UMR 5152 du
  CNRS,  Universit\'e Paul Sabatier, 118, route de
  Narbonne, 31062 Toulouse, FRANCE} 

\begin{abstract}We introduce an interference measure which allows to
  quantify the amount of interference present in any physical process that
  maps an initial density matrix to a final density matrix. In particular, 
  the interference measure enables one to monitor the amount of interference generated
  in each step of a quantum algorithm. We show that a Hadamard gate acting
  on a single qubit is a 
  basic building block for interference generation and realizes one  bit of
  interference, an ``i-bit''. We  use the interference measure 
  to quantify interference for various examples, including Grover's
  search algorithm and Shor's factorization algorithm. We distinguish between ``potentially
  available'' and ``actually used'' interference, and show that 
  for both algorithms the potentially available interference is
  exponentially large. However, 
  the amount of interference actually used in Grover's algorithm is
  only about 3 i-bits and asymptotically independent of the number of qubits,
  while Shor's algorithm indeed uses an exponential amount of interference.
\end{abstract}
\maketitle
\section{Introduction}
Entanglement and interference are believed to be key
ingredients that distinguish quantum from classical
information processing \cite{Bennett00}. Indeed it has been shown that large
amounts of entanglement must necessarily be generated in a quantum algorithm
that offers an exponential speed-up over classical computation
\cite{Jozsa03}. 
Tremendous effort has been spent to develop methods to detect 
and quantify entanglement in a given quantum state (see
\cite{Bruss01,DeSLS05} for recent reviews). On the other hand very little has
been done to quantify ``interference''. It seems therefore well worthwhile
to analyze the role of interference in quantum algorithms in more detail.

While entanglement plays an important role in many quantum information
tasks, like for example quantum teleportation \cite{Bennett93}, quantum key distribution
\cite{BennettB84}, or super--dense coding \cite{BennettW92}, large amounts of
entanglement may not be the only requirement to get a speed up with a quantum
algorithm \cite{Jozsa03,CleveEMM98}.  
Classical analogues of entanglement exist \cite{Collins02}, for 
example in the context of the propagation of classical phase space density
through Liouville's equation \cite{Lakshminarayan01}. Even without talking
about dynamics, a formal analogy of pure state quantum
entanglement can be easily defined in the classical domain, if we
replace wave functions in the definition of quantum entanglement by
probability distributions. Arbitrary large amounts 
of ``classical entanglement'' of many-particle probability distributions
(corresponding to classical correlations) can
thus be created by a classical stochastic computer. 
However, only linear combinations of
probability distributions with real positive
coefficients are allowed, and this makes it impossible to efficiently realize
Fourier-transforms with high contrast, a decisive ingredient in many quantum
algorithms (see e.g. \cite{Shor94,AbramsL99,GeorgeotS01}).  In general it  
seems that interference 
between many computational paths plays an important role in quantum computation. 
It is unclear, however,  how much interference, if any, is needed for a given quantum
speed-up. 

As a first step in the direction of answering this question, we
introduce here a general interference measure that applies to 
any physical process which maps an initial density matrix to a final density
matrix. It can be used in particular for any quantum algorithm, measurement
processes, or classical communication. We use the interference measure to
quantify interference in various steps of the well--known quantum algorithms
of Shor \cite{Shor94} and Grover \cite{Grover97} as
well as other physical examples, including photons passing through a
beam-splitter or through a Mach-Zehnder interferometer, decoherence through bit-flip and phase
errors, and quantum teleportation.

\section{The Essence of Interference} 
The familiar example of a double slit
experiment, where waves
with different wave-vectors are superimposed at the 
slits, propagate and interfere to generate an interference pattern,
i.e.~a certain probability distribution on
the screen in a position basis, contains the basics of any interference
experiment: 
\begin{enumerate}
\item\label{coh} Interference needs coherence.  An interference
  measure has to 
  distinguish between coherent and incoherent propagation. This means
  first of all that interference is a property of a
  propagator of states, not of a state itself, in contrast to
  entanglement.  
\item Coherent propagation alone does not amount to interference, however. For
example a quantum gate which implements just the unitary identity operation
is completely coherent (a pure state remains a 
pure state), whereas there is no interference at all: the probability
for each final basis state depends only on the initial
 amplitude of the same basis state. Similarly, there is no
 interference for a quantum gate that just permutes in-coming amplitudes. 
Evidently, at
least two states have to be coherently superposed. Generally speaking, the
amount of interference 
should be linked to how many different initial state amplitudes
contribute to each final state amplitude, and to 
what extent. An interference measure should be maximized if each
basis state as input state produces an ``equipartitioned'' output state,
i.e.~a state with the same 
absolute probability amplitude for each basis state.
\item Interference is basis dependent. Indeed, no interference
  pattern at all, but a probability distribution just reflecting the
  initial amplitudes would be observed in the double slit experiment
  described above,
  if the output were observed in the momentum basis. Also any unitary
  evolution is given by a unitary matrix $U$ defined
  in a specific basis (for example the computational basis for a quantum
  algorithm). This matrix can 
  always be diagonalized, and in the new basis the propagator then
  just corresponds to a multiplication with a phase factor for each
  input basis state, which does clearly not amount to interference.  
\end{enumerate}

\section{The interference measure}
\subsection{General formulation}
We start by defining a measure of coherence for a general propagator $P$ of a
density matrix, $\rho'=P\rho$, where 
$P$ is a super-operator specified in the computational basis, such that, written in components,
\begin{equation} \label{rhoij}
\rho_{ij}'=\sum_{k,l}P_{ij,kl}\rho_{kl}\,.
\end{equation}
Further below we will adapt this formalism to the ``operator sum formalism'' more
familiar in the context of quantum computation \cite{Nielsen00}.

Consider a black box which maps an initial density matrix to a 
final probability distribution, as is normally the case for a quantum
  computer if the read-out process is considered part of the
  algorithm. As long as 
one regards a unique input  state (pure or mixed),
one cannot distinguish between coherent and incoherent
propagation. Indeed, the black box might forget about any initial phase
information, keep just the initial probabilities and then use a stochastic
matrix which maps them to the probability distribution of 
the desired final state. The final
probabilities are then invariant under arbitrary phase rotations of the
initial amplitudes. To distinguish between coherent and
  incoherent propagation,  an interference measure must
therefore quantify the dependence of the final probabilities on the initial
phases --- a strategy which is in fact
often employed experimentally to show coherence.

Let us therefore start with a pure
initial state,  $\rho=|\psi\rangle\langle \psi|$, with
$|\psi\rangle= \sum_{j=1}^N a_j|j\rangle$.
The amplitudes $a_j$ with phases $\varphi_j$, 
$a_j=|a_j|\re^{\ri\varphi_j}$, lead to final probabilities 
\begin{eqnarray}
p_i'=|\rho_{ii}'|&=&\sum_{j,k}P_{ii,jk}\re^{\ri(\varphi_j-
  \varphi_k)}|a_ja_k|\,.\label{pi}   
\end{eqnarray}
The dependence of each individual probability on the initial phases is
thus given by
\begin{eqnarray}
\frac{\partial p_i'}{\partial \varphi_l}&=&\ri
\sum_{k}|a_ka_l|\left(P_{ii,lk}\re^{\ri(\varphi_l-\varphi_k)}-c.c.\right)\,.
\end{eqnarray}
We define the real phase sensitivity matrix $S$ with matrix
elements $S_{il}=\partial p_i'/\partial \varphi_l$, and the 
positive semi--definite matrix $S S^T$, with $S S^T=0$ iff
$\partial p_i'/\partial \varphi_l=0$ for all $i,l=1,\ldots,N$. 
A coherence measure ${\cal C}(P)$ can be obtained by taking the
trace of $S S^T$ and averaging it over all initial phases,
\begin{eqnarray}
{\cal
  C}(P)&=&\frac{1}{(2\pi)^N}\int_0^{2\pi}d\varphi_1\ldots d\varphi_N
  \tr(SS^T) \\\label{Cini}
&=&\sum_{i,k,l,m}|a_ka_l^2a_m|\left[-\int_0^{2\pi}\frac{d\varphi_k
  d\varphi_l d\varphi_m}{(2\pi)^3}
  P_{ii,lk}P_{ii,lm}\re^{\ri(2\varphi_l-\varphi_k-\varphi_m)}
+ \int_0^{2\pi}\frac{d\varphi_k
  d\varphi_m}{(2\pi)^2}P_{ii,lk}P_{ii,ml}\re^{\ri(\varphi_m-\varphi_k)}
\right]+c.c\nonumber\\
&=&2\sum_{i,k}\sum_{l\ne k}|P_{ii,lk}a_ka_l|^2\,.
\end{eqnarray}
This measure still depends on the amplitudes of the initial
state. However, coherence should be measured for an input state with
amplitudes on all basis states. We therefore chose a ``democratic''
input state with $|a_i|=1/\sqrt{N}$ for all $i=1,\ldots,N$. We multiply with
an additional prefactor $N^2/2$ and define the coherence measure
\begin{equation} \label{IM}
\cI(P)=\sum_{i,k,l}|P_{ii,kl}|^2-\sum_{i,k}|P_{ii,kk}|^2\,.
\end{equation}
The quantity $\cI(P)$ obviously has the desired property to be zero in the
case of a classical stochastic propagation, that is for
$P_{ij,kl}=M_{ik}\delta_{ij}\delta_{kl}$, where $M_{ik}$ is a stochastic
matrix which propagates classical probabilities from initial values $p_k$ to
final values $p_i'$, $p_i'=\sum_{k}M_{ik}p_k$,
and $\delta_{ik}$ denotes the Kronecker--delta. From the definition 
it is clear that $\cI(P)$ is non--negative. 
In the case where all eigenvalues of $P$ are smaller than one, as for
example for dissipative quantum maps
an upper bound is given by $\cI(P)\le N^2$, as can be seen by using the
inequality $|| P \psi||\le || \psi||$ with
$\psi_{kl}=\delta_{k,k_0}\delta_{l,l_0}$ for all $1\le k_0,l_0\le N$. This
bound is probably not optimal, and, as
will be seen, can be improved in the case of unitary propagation. 

The propagation of mixed states is in quantum information theory generally
formulated within the operator sum formalism \cite{Nielsen00}: A set of
operators $\{E_l\}$ acts on $\rho$ according to $\rho'=\sum_l E_l\rho
E^\dagger_l=P\rho$. The $E_l$'s are known as Kraus operators \cite{Kraus83}
and obey $\sum_k E_k^\dagger E_k={\bf 1}$ for trace--preserving operations. The
connection to the 
propagator $P$ is given 
by $P_{ij,km}=\sum_l(E_l)_{ik}(E_l^*)_{jm}$, and we can reformulate the
interference measure (\ref{IM}) as   
\begin{equation} \label{IME}
\cI=\sum_{i,k,m}\left|\sum_l(E_l)_{ik}(E_l^*)_{im}\right|^2-\sum_{i,k}\left(\sum_l\left|(E_l)_{ik}\right|^2\right)^2\,.
\end{equation}
We will now show that $\cI(P)$
is in fact a good measure of interference, as it also measures the
amount of equipartition in the case of unitary propagation.

\subsection{Unitary propagation}
Coherence is, by definition, perfect, if all pure incoming states remain
pure during propagation, i.e.~also the final state can be written as
$\rho'=|\Phi\rangle\langle \Phi|$, with a state $|\Phi\rangle$ related to
$|\psi\rangle$ by a unitary transformation, $|\Phi\rangle=U|\psi\rangle$. In
this case $P$, which we shall denote by $P(U)$, has matrix elements 
$P_{ii,kl}=U_{ik}U^*_{il}$. 
Profiting
further from the fact that 
$\sum_{i,k,l=1}^N|U_{il}U_{ik}|^2=N $ due to the unitarity of
$U$, we find in this case
\begin{eqnarray}
\cI(P(U))
&=&\left(N-\sum_{i,k}|U_{ik}|^4\right)\,.\label{cohm}
\end{eqnarray}
In the unitary case the coherence measure (\ref{cohm}) has the property that
$0\le \cI(P(U)) \le N-1$. The right hand side of this inequality is easily
verified using the Cauchy--Schwarz inequality applied to all vectors
$(|U_{i1}|^2,\ldots,|U_{iN}|^2)$ ($i=1,\ldots,N$), and
$(1/\sqrt{N},\ldots,1/\sqrt{N})$, and is in fact the optimal bound, as it is
reached for $|U_{ik}|=1/\sqrt{N}$ for all $i,k$. The left hand side follows 
from the
positivity of $SS^T$, or explicitly from $\sum_{i,k}|U_{ik}|^4\le
  \sum_{i,k}|U_{ik}|^2=\sum_k 1=N$, where we have used that $0\le
  |U_{ik}|^2\le 1$ due to $\sum_i|U_{ik}|^2=1$. 

The term $\sum_{ik}|U_{ik}|^4$ is nothing but the inverse participation
ratio (IPR) of a column $k$ of the unitary matrix summed over all columns. 
The IPR is a well--known measure of ``equipartition'' of a
wave-function, used extensively in solid state physics in order to
measure localization \cite{Wegner80}. A column of $U$ with 
an amplitude on a single basis state, i.e.~$U_{ik}=\delta_{ii_k}$ for column
$k$ and some index $i_k$ 
gives $\sum_{i}|U_{ik}|^4=1$. If {\em all}  columns have an entry on just a
single 
basis state we get therefore $\sum_{ik}|U_{ik}|^4=N$.  

Thus, if $U_{ik}=\delta_{iP_i(k)}$, where $P_i(k)$ is a permutation, we have
$\cI(P(U))=0$. This reflects just the fact that this kind of coherence is useless
--- 
the final probability distribution does not depend at all on the
initial phases and for all possible input states, the same output
could be obtained with a propagation of probabilities only. 

On the other hand, perfectly ``equipartitioned'' output states for each
computational basis state used as input, $|U_{ik}|=1/\sqrt{N}$ for all $i,k$, give
$\sum_{ik}|U_{ik}|^4=1$. Therefore, $\cI$ measures for unitary propagation
also the amount of equipartition, where $\cI(P(U))$  varies between 0 for a mere
permutation of all input states and $N-1$ if all output states corresponding
to computational basis input states are perfectly equipartitioned. {\em We
therefore define ${\cal I}(P)$ in eq.(\ref{IM}) as the measure of
interference in the 
propagator $P$}. Note that this definition  is very general, and might be
applied to any physical situation where (possibly mixed) input states are
propagated to (possibly mixed) output states.

\subsection{Properties of the interference measure}
\subsubsection{Invariances}
As eq.(\ref{IM}) contains a double and triple sum over all computational basis
states, it is obvious that $\cI$ is invariant under a permutation of the
computational basis states. This goes hand in hand with the observation that
a pure permutation of the computational basis states does not generate any interference.
It is also obvious that $\cI$ is invariant under arbitrary phase changes of
any matrix element $P_{ij,kl}$. For unitary propagation, the inverse
propagation $U^\dagger$ leads to the same amount of interference,
$\cI(P(U^\dagger))=\cI(P(U))$. 
\subsubsection{Equipartition in a Sub-Space}
Consider a unitary matrix with $|U_{ik}|=1/\sqrt{M}$ for $1\le i,k\le M$
for some integer $M\le N$, and $|U_{ik}|=\delta_{ik}$ for $i>M$ or
$k>M$. Straightforward evaluation of $\cI(P(U))$ from eq.(\ref{cohm}) gives
$\cI(P(U))=M-1$. Thus, $\cI(P(U))$ increases indeed linearly with the number of
coherently superposed states. 
\subsubsection{Adding auxiliary qubits}
Another situation, encountered often in quantum computing, is the addition
of auxiliary qubits, or in general a Hilbert space of dimension $M$
connected by a tensor 
product to the original Hilbert space. As long as one acts on the
auxiliary qubits only with the identity operation, their only effect is to
increase the number of coherently superposed states by a factor $M$. One
should therefore expect the amount of interference of $\tilde{U}=U\otimes
  {\bf 1}_M$ 
to be larger by a factor $M$ compared to ${\cal I}(P(U))$. This is indeed the
case. To see this, define the matrix elements of $\tilde{U}$ as
$\tilde{U}_{nmkl}=U_{nk}{\bf 1}_{ml}=U_{nk}\delta_{ml}$. Then we have from
eq.(\ref{cohm}),
$\cI(P(\tilde{U}))=NM-\sum_{n,m,k,l}|U_{nk}|^4\delta_{ml}=M(N-\sum_{n,k}|U_{nk}|^4)=M\cI(P(U))$. 
The corresponding calculation can be done for the general non--unitary
propagation of a density matrix 
$\rho=\rho^A\otimes \rho^B$, such that
$\rho'=\tilde{P}\rho=(P\rho^A)\otimes\rho^B$, and the result is the same:
$\cI(\tilde{P})=M\cI(P)$. 
\subsubsection{The ``i-bit'' and the Hadamard gate}\label{sec.ibit}
The linear dependence of the interference measure on the number of
coherently superposed states makes it possible to define a unit of
interference. We can define the number $n_I$ of ``interference bits'' (or
``i-bits'' for short) that measures the (logarithmic) amount of interference
in a propagator $P$ as 
\begin{equation} \label{nI}
n_I=\log_2(\cI(P)+1)\,.
\end{equation}
 As an immediate consequence we find that a Hadamard gate,
 $H_{ij}=(-1)^{ij}/\sqrt{2}$ for $i,j \in {0,1}$, 
generates one bit of interference as $\cI(P(H))=1$, and this is the maximum
 possible amount of interference for the unitary propagation of a
single qubit. Moreover, it is easy
to show that for a tensor product of $n$ Hadamard gates acting in parallel
 on $n$ qubits, also called the 
Walsh--Hadamard transform, $W=H\otimes
  H\otimes\ldots\otimes H$,   one obtains $n$
i-bits, $\cI(P(W))=2^n-1$, as the amplitude of each matrix element of
  the tensor product has an absolute value $1/2^{n/2}$. One may thus
  consider  
the Hadamard gate as a ``currency'' of interference in a quantum algorithm,
much as 
a singlet measures the amount of entanglement in a bipartite quantum
state. Note, however, that not each Hadamard gate in a quantum algorithm
adds an i-bit to the total amount of interference.
If $p$ Hadamard gates act on $p$ different qubits out of a
 total number  of $n$ qubits, the interference is $\cI=2^n-2^{n-p}$.

\subsection{Potentially available interference versus actually used interference}
Another reason why a Hadamard gate does not necessarily add an i-bit of
interference to a quantum algorithm lies in the fact that the amount of
interference added in a given 
step of the algorithm depends on the transformation built in all previous
steps. This is easily seen from 
the example of two Hadamard gates in series, acting on
a single qubit, $H^2={\bf I}_2$, such that $\cI(P(H^2))=0$. We will call
``accumulated interference'' the total interference 
$\cI(P_m\ldots P_1)$ of a sequence of transformations $P_m\ldots P_1$, which
is in general very different from the sum of interferences of each step
$P_i$. In principle one can calculate the accumulated interference for an
arbitrary part of any quantum algorithm, but most interesting is in general
the accumulated interference of the entire algorithm. 

It turns out that many quantum algorithms generate a lot of interference right
at the beginning, including Shor's and Grover's algorithms (see below), as
they
start out with the Walsh--Hadamard transform on many qubits.
In the end the desired information (e.g. the period of a function or the label
of a searched state) has to be extracted, which is done by a {\em reduction}
of the accumulated interference, as the probability distribution gets
concentrated on only 
very few computational states. The
Fourier transform, which taken by itself would add even
more interference, performs this task  in Shor's algorithm. 

However, not all
quantum algorithms actually {\em use} all the interference generated. A tremendous
``waste'' of interference can arise when 
only a single column of
the unitary matrix corresponding to a single initial state (e.g.~
$|0\ldots0\rangle$) is used, whereas it is completely irrelevant what happens
in the other columns. One might be tempted to calculate the interference
just corresponding to that initial state. Formally this can be done by
including a 
projection onto the state $|0\ldots 0\rangle$ in the algorithm, but it is
easily shown that then the accumulated interference of this projection
combined with whatever unitary
operation follows vanishes. All coherence is destroyed in the sense that the
final probability distribution is independent of the single remaining
initial phase. An alternative approach is to  
calculate the accumulated interference for the remaining algorithm
$\tilde{U}$ after the initial Walsh--Hadamard transform, i.e.~once 
amplitudes from all computational states are available in the
superposition. This amounts to changing the initial state. Then all columns
of $\tilde{U}$ do contribute, and as a 
consequence $\cI(P(\tilde{U}))$  is a
more realistic measure of the interference actually used.  
The interference measures corresponding to $U$ and $\tilde{U}$
therefore give
complementary information (total ``potentially available'' interference in the algorithm
and ``actually used'' interference, respectively) and will be calculated
below for Grover's and Shor's algorithms.

\section{Applications}
\subsection{Beam splitter}
\begin{figure}
\includegraphics{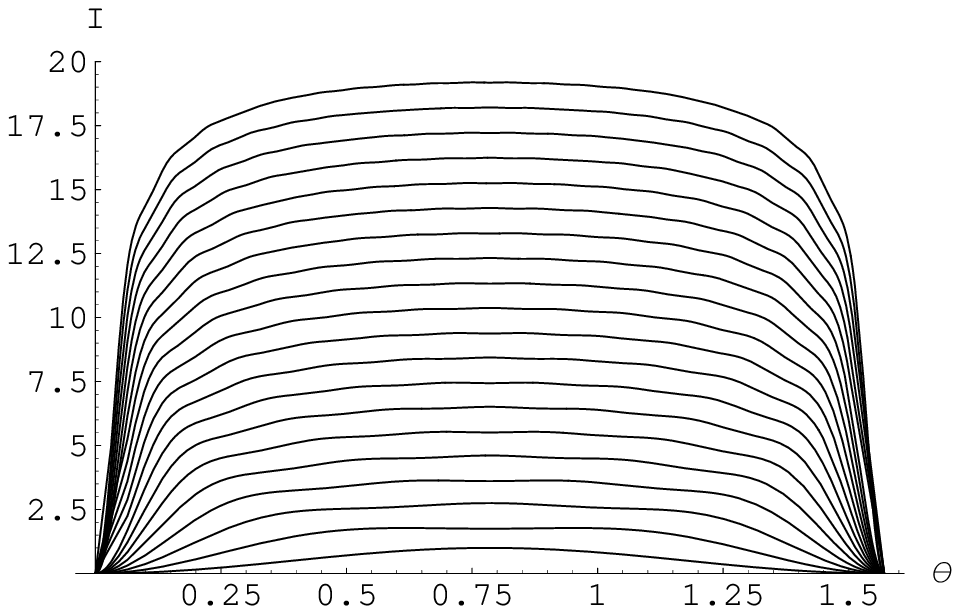} 
\includegraphics{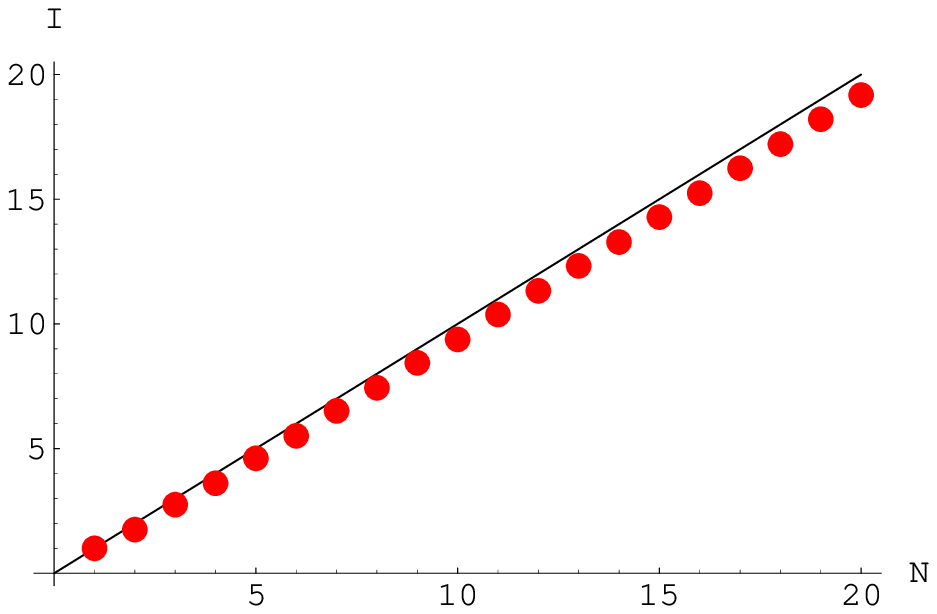}  
\caption{(Color online) Amount of interference $\cI$ generated by a beam
  splitter as function of 
  the angle $\theta$ and $N=1,\ldots 20$ photons (top). The amount of
  entanglement at $\theta=\pi/4$ increases approximately linearly with $N$
  (bottom). The full line is $\cI=N$.}\label{fig.BS}       
\end{figure}

A beam splitter transforms two photon modes with annihilation (creation)
operators $a,b$ ($a^\dagger,b^\dagger$), respectively, according to the unitary
transformation $U_{\rm BS}=\exp(\theta(a^\dagger b- ab^\dagger))$
\cite{Nielsen00}. The action of $\Ubs$ on a
state $|nm\rangle$ with $m$ photons in mode $a$ and $n$ photons in mode $b$
can be easily derived with the help of the relations $\Ubs a \Ubs^\dagger=a
\cos\theta+b\sin\theta$ and $\Ubs b \Ubs^\dagger=b
\cos\theta-a\sin\theta$. We exploit the fact that the total 
number $N$ of photons in the two modes is conserved to write the matrix
elements of $\Ubs$ as     
\begin{eqnarray} \label{UBS}
(\Ubs^{(N)})_{im}&\equiv& \langle i\,\, N-i |\Ubs| m\,\,
  N-m\rangle\nonumber\\
&=&\sqrt{\frac{i!(N-i)!}{m!(N-m)!}}
  \sum_{l={\rm Max}(m-i,0)}^{{\rm Min}(N-i,m)}{m\choose l}{N-m\choose
  N-i-l}(-1)^l (\cos\theta)^{m+N-i-2l}(\sin\theta)^{i-m+2l}\,.
\end{eqnarray}
For example, the dual-rail representation \cite{Nielsen00} with logical
states $|01\rangle$ 
and $|10\rangle$ leads to the $2\times 2$ rotation matrix 
\begin{equation} \label{UBS1}
\Ubs^{(1)}=\left(
\begin{array}{cc}
\cos \theta&-\sin\theta\\
\sin\theta& \cos\theta
\end{array}
\right)\,,
\end{equation}
with the amount of interference
$\cI(P(\Ubs^{(1)}))=2(1-\cos^4\theta-\sin^4\theta)$. Thus a maximum
interference of 1
i-bit can be achieved for $\theta=\pi/4$, in which case the beam splitter
realizes indeed just a Hadamard gate, up to phase shifts. The function
$\cI(P(\Ubs^{(N)}(\theta)))$ appears to be 
periodic with period $\pi/2$ for all $N$. The maximum amount of interference
increases approximatively
linearly with $N$ and is reached for sufficiently large $N$ at
$\theta=\pi/4$. With increasing $N$, the 
maximum is reached more and more rapidly as function
of $\theta$, and $\cI$ remains almost flat in a broader and broader interval
around the maximum (see figure \ref{fig.BS}).

\subsection{Mach-Zehnder interferometer}
\begin{figure}
\includegraphics{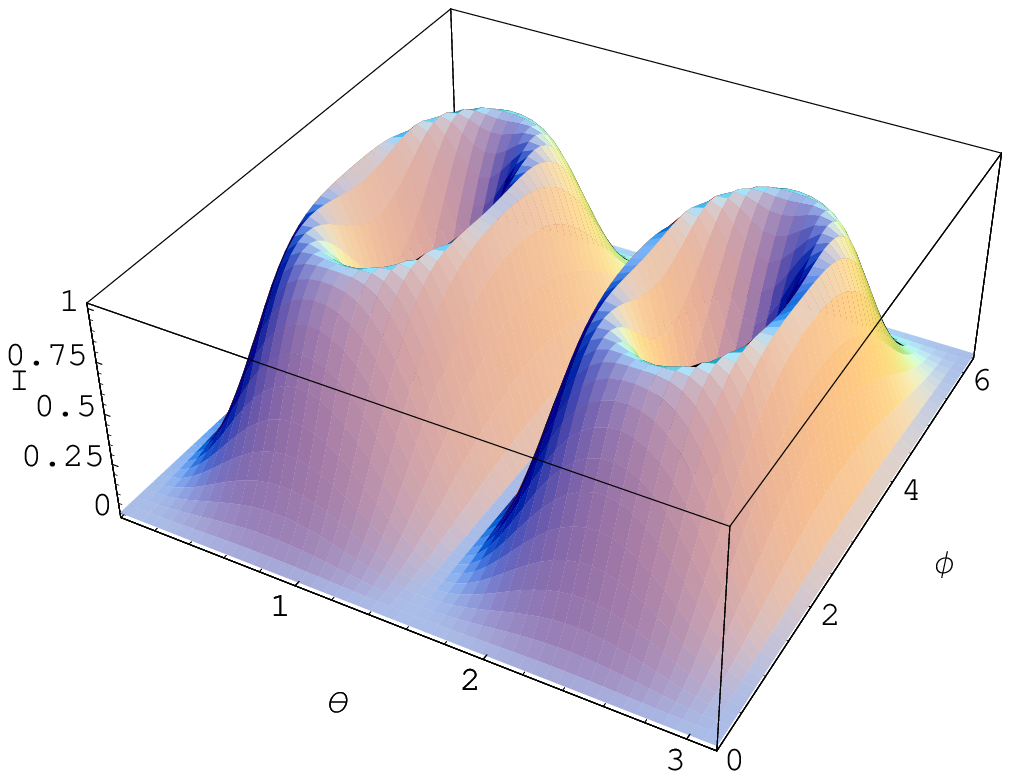} 
\includegraphics{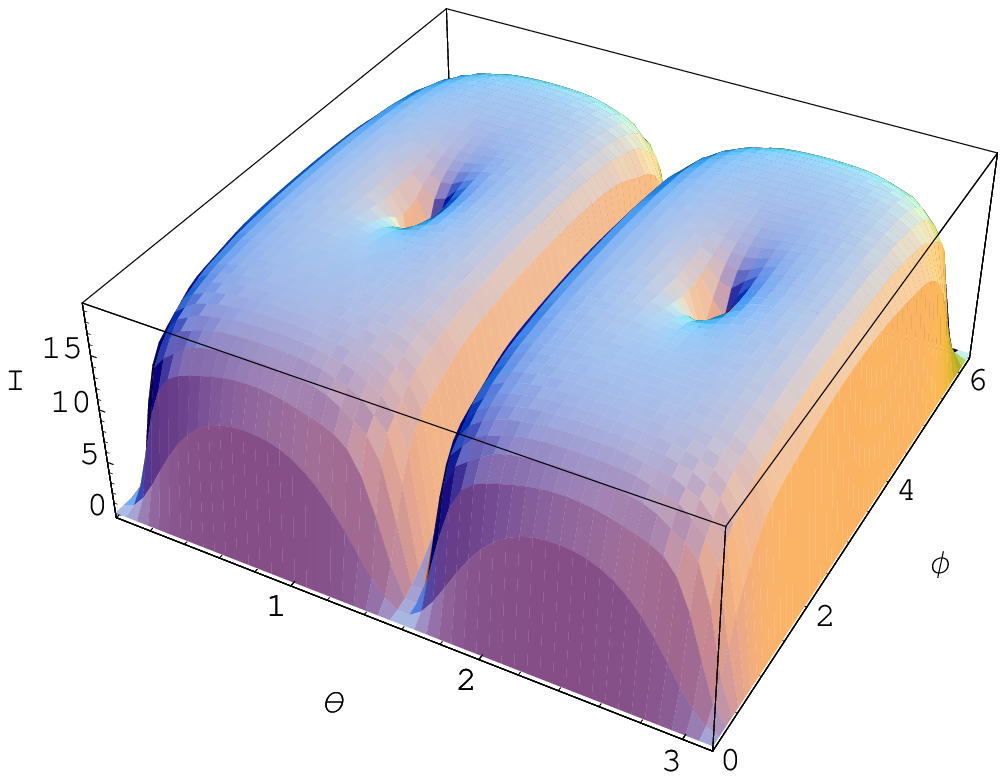} 
\caption{(Color online) Accumulated interference generated in a Mach-Zehnder interferometer as
  function of 
  the angles $\theta$ and $\phi$ for 1 photon (top) and 20
  photons (bottom). Note the different scale.}\label{fig.MZ}       
\end{figure}

A Mach-Zehnder interferometer consists of two beam splitters in series, with
an additional phase shifter in one of the two arms \cite{Nielsen00}. The second
beam splitter is inverted relative to the first one, such that the total
unitary propagation for $N$ photons is given by
\begin{equation} \label{UMZ}
U_{\rm MZ}^{(N)}=U_{\rm BS}^{(N)}U_{\rm P} U_{\rm
  BS}^{(N)\dagger}\,. 
\end{equation}
The phase shifter acts on a $m$ photon state in mode $a$ as $U_{\rm
  P}|nm\rangle= \re^{\ri \phi m}|nm\rangle$.  
  The interference generated by the Mach-Zehnder interferometer is
  periodic in $\theta$ and
  $\phi$ with periods $\pi/2$ and $2\pi$, respectively
  (Figure \ref{fig.MZ}). Again, 
  $\cI(P(U_{\rm MZ}^{(N)}))$ becomes more and more flat around the maximum with
  increasing $N$. The settings $\theta=\pi/2$, or $\theta=\pi/4$ or $3\pi/4$
  and $\phi=\pi$,  lead to a minimum of $\cI$ in the range of $N$
  investigated ($N=1,\ldots,20$). Finally, $\phi=0$ never leads to any interference,
  as in this case the interferometer just performs the identity operation. 

\subsection{Decoherence: bit flip errors and phase errors} 
The error operators for a bit flip error in a single qubit are given by
$E_0=\sqrt{p}{\bf 1}_2$, $E_1=\sqrt{1-p}X$, where $X$ is the Pauli
$\sigma_x$ matrix and $p$ the probability for no error to
occur. Similarly, the error operators for a phase error in a single qubit
read $F_0=\sqrt{p}{\bf 1}_2$, $F_1=\sqrt{1-p}Z$, where $Z$ is the Pauli
$\sigma_Z$ matrix. One easily shows using eq.(\ref{IME}) that the amount of
interference vanishes for both types of errors, as it should of course be,
due to the fact that these are purely decohering processes. More interesting
is the situation where we first apply a Hadamard gate and then the
errors. That is, the Kraus operators are now ${E_0H,E_1H}$ for a bit flip
error after the Hadamard gate, and ${F_0H,F_1H}$ for a phase
error after the Hadamard gate. In the first case, we obtain
$\cI=(1-2p)^2$. As it should be, the interference is completely conserved,
$\cI=1$, for $p=0$ or $p=1$, i.e.~when bit flips either never occur or do
occur with certainty. Even in the latter case the interference is perfect,
as the bit flip just corresponds to a unitary permutation. On the other
hand, in the case of a phase flip after a Hadamard gate, interference is
{\em always} conserved, i.e.~$\cI=1$ independent of $p$. While this might be
surprising at first, it is in fact a remarkable property of $\cI$ to
detect that the phase error does not modify at all an interference pattern
obtained by coherently superposing $|0\rangle$ and $|1\rangle$. To see this,
we start with the initial state
$(\re^{\ri\varphi}|0\rangle+|1\rangle)/\sqrt{2}$. After the Hadamard gate
this state becomes $(\re^{\ri
  \varphi}(|0\rangle+|1\rangle)+(|0\rangle-|1\rangle)/2$. With probability
$p$ the state remains the same, while with probability $1-p$ a phase error
flips the sign in front of the $|1\rangle $ states. One obtains thus the
final density matrix $\rho_f=\cos^2(\varphi/2)|0\rangle\langle
0|+\sin^2(\varphi/2)|1\rangle\langle
1|+(1/2-p)\ri\sin\varphi(|1\rangle\langle 0|-|0\rangle\langle 1|)$. Clearly,
the probabilities to find 0 or 1 in the final state are unaffected by the
phase error, and one gets a perfect interference picture for any value of
$p$, as the phase errors from the bra and the ket cancel. Note that by just
looking at the off-diagonal matrix elements of $\rho_f$ 
(as is often done to estimate the amount of coherence in a {\em state})
one concludes that complete decoherence and the reduction to a
classical mixture occurs at $p=1/2$, whereas for $p=0$ or $p=1$ a completely
coherent final state is retained. This is, however, irrelevant
for the success of the interference experiment, and correctly detected by
our interference measure.

More generally, one might consider the action of
the phase error as a measurement of the interference pattern. Indeed, $Z$ is
diagonal in the computational basis, which thus constitutes the ``pointer
basis'' of $Z$ \cite{Zurek81}. The measurement reveals the interference
pattern, but does not destroy it. It is therefore reasonable to demand that
an interference measure be conserved during the measurement process. 
The fact that interference can be conserved during measurements is crucial
also for quantum teleportation.

If a phase error occurs after the phase shifter in the Mach-Zehnder
interferometer, the interference is reduced. For the example of one photon
the  Kraus operators are given by $G_0=\sqrt{p} U_{\rm BS}^{(1)}U_{\rm P} U_{\rm 
  BS}^{(1)\dagger}$ and $G_1= \sqrt{1-p} U_{\rm BS}^{(1)}ZU_{\rm P} U_{\rm 
  BS}^{(1)\dagger}$. The general expression for the interference is somewhat
cumbersome, but simplifies for beam splitters with $\theta=\pi/4$ to
$\cI=(\sin\varphi(1-2p))^2$. So the interference in the Mach-Zehnder
interferometer is clearly completely destroyed for $p=1/2$, as is to be expected.

\subsection{Quantum Teleportation}
Alice can teleport an unknown quantum state of a qubit 1 to
Bob, by doing the 
following \cite{Bennett93}: She first prepares two auxiliary qubits (2,3) in
the  
Bell state 
$(|00\rangle+|11\rangle)/\sqrt{2}$, and 
sends qubit 3 to Bob. We will consider the Bell state preparation as part of
the 
protocol, as it will become clear that already here 
a certain amount of interference is used \footnote{One might of course
argue that if  Alice has to send a qubit, she might as well just send her
own qubit 
1. Normally the Bell pair would in fact be prepared by an independent
source, which sends one qubit to Alice and the other to Bob. Nevertheless,
it is instructive to consider the Bell pair creation process as part of the
protocol.}. Starting from 
state $|00\rangle$ of qubits (2,3), Alice can prepare the Bell state by
applying $H$ to 
qubit 3 ($H_3$, step 1) and then a CNOT with 3 as control and 2  as
target ($C_{32}$) to qubits 
(2,3) (step 2). Next 
Alice applies a CNOT with 1 as control and 2  as
target
($C_{12}$) to qubits (1,2) (step 
3) and a Hadamard to qubit 1 ($H_1$, step 4). She then measures qubits (1,2)
in the 
computational basis (step 5), and sends the result ${m_1,m_2}$ (where
$m_i=$ 0 or 1) to 
Bob through a classical 
channel. Bob performs the unitary operation $Z^{m_1}X^{m_2}$ (step 6). The
total 
propagation of the density matrix thus reads   
\begin{equation} \label{qtp}
\rho'=\sum_{i,j=0,1}Z^{j}X^{i}E_{ij}H_1C_{12}C_{32}H_3\rho
\left(Z^{j}X^{i}E_{ij}H_1C_{12}C_{32}H_3\right)^\dagger  \,,
\end{equation}
where the $E_{ij}$ are projectors, $E_{ij}={\bf 1}\otimes |ij\rangle\langle
ij|$, and the qubits are numbered from right to left. 
\begin{figure}
\includegraphics{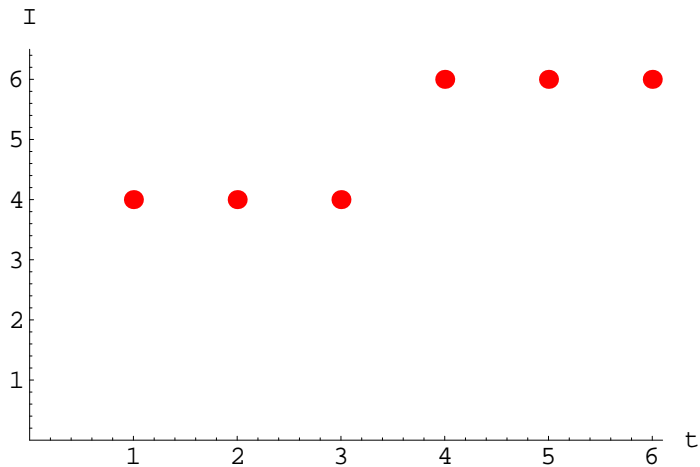} 
\caption{(Color online) Accumulated interference  during the quantum teleportation protocol
  \cite{Bennett93}. On the t-axis is the step number (see
  text). The interference is plotted for the entire propagator up to and
  including step 
  number $t$. Interference only arises during the application of the two
  Hadamard  
  gates and is conserved during all other steps, including the measurement
  (step 5). A maximum amount of interference of about 2.58 i-bits ($\cI=6$) is
  generated.}\label{fig.qtp}      
\end{figure}
Figure \ref{fig.qtp} shows how the amount of interference used in this
protocol evolves step by step. Obviously already the creation of the initial
entangled state uses interference, through the application of the Hadamard
gate in qubit 3 (step 1). Qubits 1 and 2 are left alone, therefore the total
amount of interference is $\cI (P(H\otimes {\bf 1}_4))=4$. 
In fact, interference
is {\em only} generated by 
Hadamard gates in this algorithm, and the largest amount of interference
reached is $\cI=6$, or almost three i-bits. Remarkably, the interference is
not destroyed in the measurement process.

\subsection{Shor's algorithm}

Shor's algorithm \cite{Shor94} enables one to factor a large integer number
$R$  into primes
using a polynomial number of operations. 
The initial state is prepared in two registers of size $2L$ and $L$,
where $L=[\log_2 R]+1$ ($[.]$ denotes the integer value), with a total
number of qubits $n=3L$. 
The algorithm can be
decomposed into three parts. First, the state
$|0\ldots 0\rangle|0\ldots 01\rangle$ is transformed into 
$N^{-1/2}\sum_{t=0}^{N-1}|x \rangle|1\rangle$ (where $N=2^{2L}$
equals the dimension of the Hilbert space of $2L$ qubits)
by the use of $2L$ Hadamard gates applied to each qubit separately.
As mentioned in section \ref{sec.ibit}, this part generates the amount of
interference,
$\cI= 2^{n}-2^L$.
This value of $\cI$ corresponds to the maximum value of $\cI$ for
transformations of the first register alone. On the other hand, no entanglement
is created since the state remains factorizable.
In a second part, the state
is transformed in $O(n^3)$ operations into
$N^{-1/2}\sum_{t=0}^{N-1}|x \rangle|f(x)\rangle$ where
$f(x)=a^x \; (\mbox{mod} \;R)\;$.
This part can be viewed as a permutation matrix: each
line
and each column has only one nonzero entry. Therefore the interference
measure for this transformation alone gives zero. 
On the other hand, it creates entanglement.
The third part consists of a Quantum Fourier Transform (QFT) on the first
register only which allows one to find
the period of the function $f$. The corresponding operator 
on the first register can be written
as a matrix whose entries are all of absolute value
$1/\sqrt{N}$.
As such, the QFT alone maximizes the interference measure (\ref{IM})
on the first register, corresponding to $\cI= 2^{n}-2^L$ ,
in the same way as the
Walsh-Hadamard transform. In contrast
with the latter,
the QFT contains two-qubit gates and creates some
entanglement. 
We note that nevertheless 
numerical simulations for small number of qubits 
seem to indicate that most of the entanglement
is created during the second phase
where no interference is built up or used \cite{KendonM04}.

It is interesting to note that various versions of the factorization
algorithm have been proposed which aim at minimizing the number of gates
and/or
qubits \cite{Zalka98,Beauregard03}. They often use 
quantum Fourier transforms to
gain efficiency
in the process $N^{-1/2}\sum_{t=0}^{N-1}|x \rangle|0\ldots 01\rangle$
$\rightarrow N^{-1/2}\sum_{t=0}^{N-1}|x \rangle|f(x)\rangle$,
for example using instances of the Sch\"onhage-Strassen algorithm
which is more efficient than usual multiplication for very large numbers.
Thus these
methods generate some interference in the second part
of the algorithm, where usually only entanglement is produced. In some sense,
this
``acceleration'' of the factorization algorithm is made by
trading interference and entanglement.

\begin{figure}
\includegraphics[width=.5\linewidth]{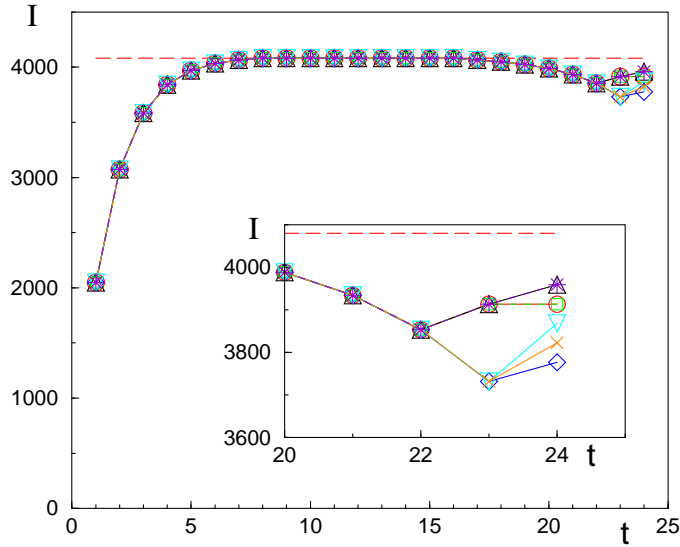} 
\caption{(Color online) Accumulated interference generated during Shor's algorithm
  for factorization of $R=15$ ($L=[\log_2 R]+1=4$ 
and in total
$n=3L=12$ qubits). On the t-axis is the
  step number (see 
  text). 
The interference is plotted for the entire propagator up to and
  including step 
  number $t$. Values of $a$ are $a=13$ (circles, full red line),
$a=7$ (squares, dashed green line), $a=11$ (diamonds, full blue line),
$a=8$ (triangles up, full black line), $a=14$ (triangles down, full cyan line),
$a=4$ (crosses, full orange line), $a=2$ (stars, dashed violet line). Data
 for different values
of $a$ differ only for the last two time steps. Data
from $a=13$ and $a=7$ are the same, as are data from $a=8$ and $a=2$.
The horizontal dashed red line is the maximum possible value of interference 
for an untouched second register, $\cI=2^n-2^L=4080$.
The inset shows the same curves for the last steps on a different scale.
Lines are there to guide the eye only.
Massive interference $\cI=2^n-2^L$ (or almost $n$ i-bits)
 is
  generated during the 
  application of the Walsh-Hadamard  
  gate (part 1, first eight points). Interference is unchanged in part 2
(next eight points), and decreases during
  the last part (QFT) (last eight points).  
  }\label{fig.shor}      
\end{figure}
\begin{figure}
\includegraphics[width=.5\linewidth]{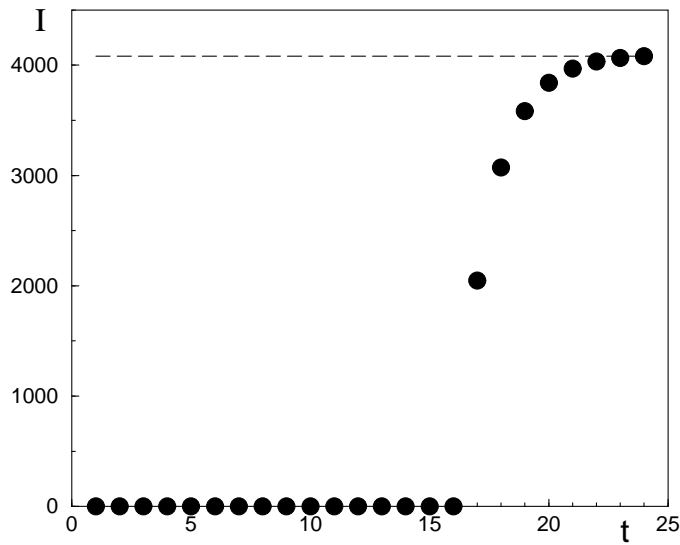} 
\caption{Accumulated interference up to and
  including step 
  number $t$ generated during Shor's algorithm,
  excluding the initial Walsh--Hadamard transform
for factorization of $R=15$ ($n=12$ qubits). The 
  step number on the t-axis is the same as in  figure \ref{fig.shor} 
but shifted by $8$.
Horizontal dashed red line is maximum possible value of interference 
 for transformations of the first register alone $\cI=4080$.
   The accumulated
  interference is changed only during the final quantum Fourier transform,
   and does not depend on the value of $a$. The
  actually used interference shown here is comparable
with the potentially
  available interference shown in figure \ref{fig.shor}.}\label{fig.shorbis}      
\end{figure}

Figure \ref{fig.shor} shows the accumulated interference for the factorization
of $R=15$.  Shor's algorithm requires $a$ to be chosen coprime to $R$.
We therefore show data for the seven possible values $a=2,4,7,8,11,13,14$.  
The initial Walsh-Hadamard transform corresponds to the first eight 
value of time 
(the interference measure is calculated after each Hadamard gate).
As explained in \ref{sec.ibit}, after $k$ Hadamard gates the value
of $\cI$ is  $\cI=2^n-2^{n-k}$.  The whole transform 
generates an exponential
amount of interference $\cI= 2^{n}-2^L=4080$,
which corresponds to the maximum value possible for a transformation
of the first register alone.  

The next eight time values correspond to the
construction of $f(x)=a^x \; (\mbox{mod} \;R)\;$ in the second register.
Each time value corresponds to multiplication modulo $R$
of the second register by $a^{2^i}$, controlled by the value of the $i^{th}$
qubit of the first register.
Each operation corresponds to a permutation in the computational basis,
and does not affect the accumulated interference.

The
last part (QFT) corresponds to the last eight time values in 
figure \ref{fig.shor}. Each time step corresponds to a Hadamard gate
followed by controlled phase transformations, which together build
the QFT.  Although the QFT taken alone generates interference, its
net effect is to decrease 
the total accumulated interference by a small amount which
depends on $a$.  
This is actually reasonable, since this last part
concentrates  
probability on a certain number of states which depend on the value of $a$
and $R$.  It is interesting to note that the first six steps
give the same values of $\cI$ for all values of $a$.  Only in the last two
steps (corresponding to the least significant bits in the first register)
does one see a branching.  The first one distinguishes between the
two values of the period of $f(x)=a^x \; (\mbox{mod} \;R)\;$
(the period is $2$ for $a=4,11,14$ and $4$ for $a=2,7,8,13$),
 and the last operation
distinguishes further between values of $a$.
Data from $a=13$ and $a=7$ are the same, as are data from $a=8$ and $a=2$,
which is consistent with the fact that for these values of $a$ 
the set of values of $a^x \; (\mbox{mod} \;R)\;$ is the same.
Other values of $a$ correspond to different interference even though the
period of $f(x)$ may be the same since the total interference depends not only on the
period of $f(x)$ but also on the period of $yf(x)$.  The total interference
which was exponentially large 
after the first two parts presumably remains so after the QFT,
since for $R=15$ it goes down only by up to $7\%$.  Note that it
can go down and up in the last operations of the QFT. 

A more realistic picture of the accumulated interference ``actually used'' is
obtained by calculating the interference measure for the operator 
obtained by omitting the initial Walsh-Hadamard transform (see section III
D).  This corresponds to the same algorithm 
but with initial state $N^{-1/2}\sum_{t=0}^{N-1}|x \rangle|1\rangle$.
This is shown in figure \ref{fig.shorbis}, where
the time steps thus correspond to the last
sixteen time steps in figure \ref{fig.shor}.
The interference measure is zero after the first part 
(construction of $f(x)=a^x \; (\mbox{mod} \;R)\;$ in the second register),
which is in accordance with the fact that all eight steps
corresponding to this part can be understood as
permutations. On the contrary, the interference measure grows exponentially
in the last (QFT) part,
and reaches the maximum value $2^n-2^L$ (for transformations of the first
register only).  Thus the actually used interference is clearly exponential
for the factorization algorithm.

\subsection{Grover's algorithm}\label{sec.Grover}
Grover's algorithm $U_G$ \cite{Grover97} can find a single marked item in an
unstructured database 
of $N$ items in $O(\sqrt{N})$ quantum operations,
to be compared with $O(N)$ operations for the classical algorithm.
The algorithm starts on a system of $n$
qubits (Hilbert space of dimension $N=2^n$) with the Walsh-Hadamard
transform 
$W$, thus by building an uniform
superposition of the basis states
$N^{-1/2}\sum_{x=0}^{N-1}|x \rangle$.
Then the algorithm iterates $k$ times the same
operator $U= W R_2 W R_1$, with an optimal value $k=[\pi/(4\theta)]$ and
$\sin^2\theta=1/N$ \cite{Boyer96},
i.e. $U_G=(W R_2 W R_1)^kW$.   
As mentioned in section \ref{sec.ibit}, $W$ generates, taken for itself, the
maximum possible amount of interference ($n$ i-bits). 
The oracle $R_1$ multiplies the amplitude
of the marked item (the one searched by
the algorithm) with a factor $(-1)$, and keeps the other amplitudes unchanged.
The corresponding $N\times N$ matrix is diagonal, making evident
that its interference measure is zero. The operator $R_2$ multiplies the
amplitude of the state $|0\ldots 0\rangle$ with a factor $(-1)$,
keeping the others unchanged. By the same argument as for $R_1$,
it generates no interference. 
Therefore all interference in
the algorithm is generated by the Walsh-Hadamard transforms $W$.  It is 
interesting to note that both operators $R_1$ and
$R_2$ can be understood as multicontrolled gates which create entanglement,
whereas $W$ cannot create
entanglement as it is composed of one-qubit operations. So Grover's
algorithm alternates entanglement creation 
and interference creation during its evolution.
We note that the interference is independent of the 
label of the searched item, as the interference measure $\cI$ is invariant
under permutation of the computational basis states.  

\begin{figure}
\includegraphics{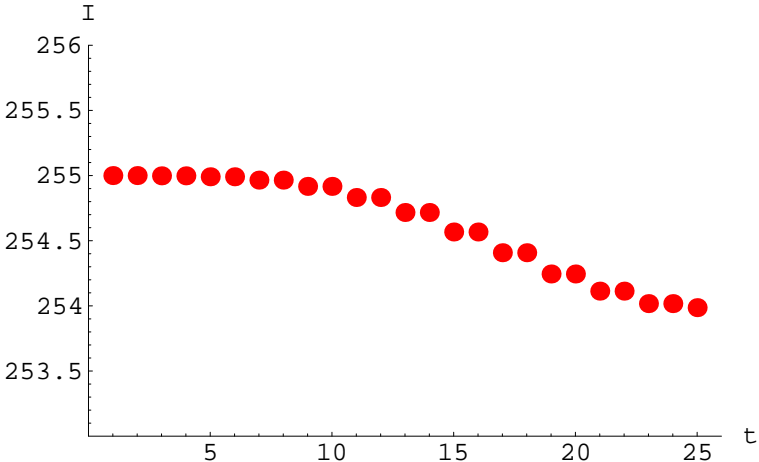} 
\caption{(Color online) Accumulated interference in Grover's algorithm
  $U_G$ \cite{Grover97} for $n=8$ qubits. On the t-axis is the
  step number (see 
  text). The interference is plotted for the entire propagator up to and
  including step 
  number $t$. Massive maximum interference $\cI_{\rm max}=2^n-1=255$ (or $n=8$ i-bits) is
  generated during the 
  application of the Walsh-Hadamard  
  gate (step 1) and the accumulated interference decreases subsequently
  during the iteration of oracle 
  and diffusion to the value $\cI\simeq 2^n-2$, at which point virtually all
  probability 
  corresponding to the initial state $|0\ldots 0\rangle$ is concentrated on
  the searched state. }\label{fig.grover}      
\end{figure}

Figure \ref{fig.grover} shows the evolution of the accumulated interference
during Grover's algorithm in $n$ qubits. In this figure, to avoid
repetition, we took the
Walsh-Hadamard transform as one single time step, contrary to the preceding
section.
Thus step 1 corresponds to the first
application of 
the Walsh-Hadamard transform, which generates the massive interference
$\cI=2^n-1$; step 2 is the first application of the oracle $R_1$, step 3 the
first application of the diffusion matrix $D=WR_2W$, step 4 the second
instance of the oracle $R_2$, and so on up to a total number of $2k+1$
steps. As expected, the oracle does not change the amount of interference,
but the diffusion matrix {\em reduces} the accumulated interference in each
step. This is crucial for the functioning of the algorithm: the
probability flow is engineered in such a way that all probability gets
concentrated on the computational basis state corresponding to the searched
item. Therefore, the accumulated interference has to decrease, as the
equipartition is decreased. 

As mentioned before, most of the interference built up during the
application of the $W$ gate is ``wasted'', as $U$ only acts on the initial
state 
$|0\ldots 0\rangle$. Indeed, no particular state is selected by any other
column of $U$.  The accumulated interference therefore reduces
only very slightly, from $\cI=2^n-1$ down to $\cI\simeq 2^n-2$. The
situation is somewhat different for Shor's algorithm where the concentration of
probability concerns many columns.  

\begin{figure}
\includegraphics{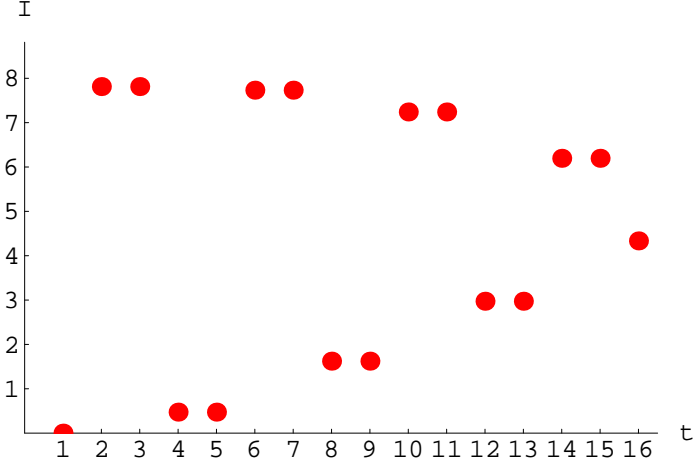} 
\caption{(Color online) Accumulated interference up to and
  including step 
  number $t$ generated during Grover's algorithm
  $\tilde{U}_G$, excluding the initial Walsh--Hadamard transform
  \cite{Grover97} for $n=7$ qubits. The 
  step number on the t-axis is shifted by one compared to figure
  \ref{fig.grover}, i.e.~step 1 is now the first application of the
  oracle $R_1$, which does not lead to interference. The accumulated
  interference is changed only during the diffusion steps $D=WR_2W$. The 
  ``actually used'' interference shown here is much smaller than the
  ``potentially 
  available'' interference shown in figure \ref{fig.grover}, and the maximum
  value $\cI\simeq 8$ after the first application of $D$ is asymptotically
  independent of the number of qubits. This is quite different from
what happens in Shor's algorithm
(compare with figure \ref{fig.shorbis}).}\label{fig.groverR}       
\end{figure}

As in the case of Shor's algorithm, it is also interesting to compute
 the actually used interference, i.e.~the interference measure for 
$\tilde{U}_G$ 
obtained by omitting the initial $W$ transform,  $\tilde{U}_G=(W
R_2 W R_1)^k$ (see section III D). Figure \ref{fig.groverR} shows that this number is much
smaller than the potentially available interference. Moreover, it is
 asymptotically  independent of the number of 
qubits. Indeed the interference for the first application of $D$ is easily
calculated analytically using the matrix elements
$D_{ij}=2/N-\delta_{ij}$ in the computational basis, and gives
$\cI(P(D))=8-24/N+{\cal O}(1/N^2)$.  Thus, Grover's algorithm actually uses
only about three i-bits, whatever the number of qubits on which it runs!
After the first application of $D$, the interference shows damped
oscillations with each application of the diffusion 
matrix, whereas 
the oracle $R_1$ leaves the interference unchanged. Note that the
oscillations would be undamped without the intermediate $R_1$ steps, as $(W
R_2 W)^2={\bf 1}$. 

Thus for the Grover algorithm
the maximum actually used
interference after the creation of the initial equipartitioned state is small
and basically independent of the number of qubits. 
This is in sharp contrast with the 
results for the Shor algorithm (preceding section) where the actually used
interference grows exponentially with the number of qubits.
One may speculate if this is the decisive
difference that leads to exponential acceleration of Shor's algorithm versus
the $\sqrt{N}$ acceleration for Grover's algorithm compared to the
corresponding best classical algorithms known. 



\section{Conclusions} We have introduced a general measure of interference,
which allows to quantify interference in any physical situation which
involves the propagation of a density matrix. We have defined a logarithmic
unit of 
interference, the ``i-bit'' and have quantified how much
interference arises in each step of the two best known
quantum algorithms as well as in other physical examples, including beam
splitter, Mach-Zehnder interferometer, and quantum 
teleportation.  A beam splitter and a Mach-Zehnder interferometer generate
an amount of interference proportional to the number of photons, and quantum
teleportation of one qubit needs about 2.58 i-bits.
Both Shor's and Grover's algorithm build up an exponential
amount of ``potentially available'' interference. However, Grover's
algorithm actually uses only about 
three i-bits, asymptotically independent of the number of qubits on which it
runs, whereas Shor's algorithm indeed uses an exponential amount of
interference. It is therefore tempting to attribute the respective
exponential versus $\sqrt{N}$ acceleration of these two algorithms to the
amount of interference actually used, but more work will be necessary in this
direction. In particular it should be very interesting to find out if
exponentially large interference is a necessary condition for an exponential
speed up of a quantum algorithm compared to its classical counterpart, or if  
the interference measure could be used to
optimize existing algorithms or to conceive
new ones.

\begin{acknowledgments}
      We thank Karol \.{Z}yczkowski and Mahn-Soo Choi for discussions
     and the IDRIS in Orsay and CALMIP in Toulouse 
for use of their computers.
This work was supported in part by the EC IST-FET project EDIQIP.
\end{acknowledgments}

\bibliography{mybibs_bt}

\end{document}